# Dual-encoded magnetization transfer and diffusion imaging and its application to tract-specific microstructure mapping.

Ilana R Leppert[1], Pietro Bontempi[2], Christopher D Rowley[1,3], Jennifer SW Campbell[1], Mark C Nelson[1,3], Simona Schiavi[2], G Bruce Pike[4], Alessandro Daducci[2], Christine L Tardif[1,3,5]

[1]McConnell Brain Imaging Centre, Montreal Neurological institute and Hospital, Montreal, QC, Canada. [2]Department of Computer Science, University of Verona, Verona, Italy. [3]Department of Neurology and Neurosurgery, McGill University, Montreal, QC, Canada. [4]Hotchkiss Brain Institute and Departments of Radiology and Clinical Neuroscience, University of Calgary, Calgary, Canada. [5]Department of Biomedical Engineering, McGill University, Montreal, QC, Canada.

Corresponding Author: Ilana R Leppert, ilana.leppert@mcgill.ca

Declarations of interest: none

## Abstract

We present a novel dual-encoded magnetization transfer (MT) and diffusion-weighted sequence and demonstrate its potential to resolve distinct properties of white matter fiber tracts at the sub-voxel level. The sequence was designed and optimized for maximal MT contrast efficiency. The resulting whole brain 2.6 mm isotropic protocol to measure tract-specific MT ratio (MTR) has a scan time under 7 minutes.  Ten healthy subjects were scanned twice to assess repeatability. Two different analysis methods were contrasted: a technique to extract tract-specific MTR using Convex Optimization Modeling for Microstructure Informed Tractography (COMMIT), a global optimization technique; and conventional MTR tractometry. The results demonstrate that the tract-specific method can reliably resolve the MT ratios of major white matter fiber pathways and is less affected by partial volume effects than conventional multi-modal tractometry. Dual-encoded MT and diffusion is expected to both increase the sensitivity to microstructure alterations of specific tracts due to disease, ageing or learning, as well as lead to weighted structural connectomes with more anatomical specificity.

# Introduction

Magnetic Resonance Imaging (MRI) offers valuable insight into the morphology, composition, and microstructural organization of white matter fibre pathways in the brain. Many MR contrasts are sensitive to tissue myelin content, including $T_1$ and $T_2$ relaxation times (review ([Does, 2018](#))) and magnetization transfer (review ([Sled, 2018](#))). These MR markers have been shown to correlate to varying degrees with myelin density derived from histology (see review ([Lazari & Lipp, 2021](#); [Mancini et al., 2020](#))). Through the combination of multiple contrast mechanisms, complementary properties of the underlying microstructure can be probed ([Callaghan et al., 2014](#); [Kolind et al., 2008](#)) and more specific microstructural indices estimated (see review in([Cercignani & Bouyagoub, 2018](#))). For instance, myelin imaging can be combined with diffusion imaging to estimate the thickness of the myelin sheath relative to axon caliber, known as the g-ratio, using a biophysical model ([Stikov et al., 2015](#)). Multiple myelin-sensitive contrasts can also be combined, for example $T_2^*$ and magnetization transfer ([Mangeat et al., 2015](#)), to help minimize the impact of confounds, such as iron content or field non-uniformity.

When different contrasts are not only combined but co-encoded, meaning they are encoded simultaneously within the same sequence, there is the potential to disentangle the signal contribution of different microstructural compartments within a voxel. For diffusion MRI, the sensitivity to both diffusivity and orientation can help dissociate MR properties of specific microstructural compartments and fiber orientations. When diffusion and relaxation are co-encoded, the relaxation times of distinct compartments and/or fiber orientations within a voxel can be measured. This is achieved by repeating the diffusion acquisition for various echo times for $T_2$, or for various repetition or inversion times for $T_1$. Different approaches to both the acquisition and the analysis of diffusion-relaxometry exist. For example, non-parametric signal inversion techniques sweep through a large range of experimental parameters and rely on limited assumptions to estimate the diffusivity and relaxation times of different compartments (Benjamini & Basser, 2016, 2020; de Almeida Martins et al., 2021; Kim et al., 2017) at the cost of being rather time consuming (Lampinen et al., 2020; Veraart et al., 2018). Others use compartment models to resolve the diffusivities and $T_2$ relaxation values of the extra- and intra-cellular compartments (Gong et al., 2020; Lampinen et al., 2020; Veraart et al., 2018).

Biophysical and signal models can also be used to resolve the distinct properties of multiple fiber populations present within a voxel. This approach relies on assumptions about compartment properties and takes advantage of the different orientations of fiber populations to disentangle tract-specific information, such as $T_1$ (De Santis et al., 2016; Leppert et al., 2021). As an alternative to voxel-wise fitting, tract-specific properties can be estimated at the streamline or bundle level, using global optimization frameworks that make the assumption that a microstructural property (e.g. the intra-cellular signal fraction per unit length) is constant along a streamline's length (Daducci et al., 2015). This global approach has been used to estimate tract-specific properties such as axon caliber and tract-specific intra-axonal $T_2$ (Barakovic, Girard, et al., 2021; Barakovic, Tax, et al., 2021), as well as myelin water fraction (Schiavi et al., 2022).

The ability to disambiguate the microstructural features of crossing white matter tracts in the brain is especially appealing in the context of tractometry studies. In conventional tractometry, quantitative or semi-quantitative MRI (qMRI) maps are projected onto reconstructed streamlines for further analysis (Bells et al., 2011; Zhang et al., 2022). In some cases, the average profile of streamlines forming a tract is computed. This has been applied to various studies of white matter development (Yeatman et al., 2012) and pathologies such as multiple sclerosis (Dayan et al., 2016; Reich et al., 2008) and stroke (Li et al., 2022). In other applications, the scalar values from qMRI maps are averaged over the whole bundle (e.g., (Correia et al., 2008; Slater et al., 2019)). Such summary qMRI measures are also used to weigh the edges of structural connectomes ([Boshkovski et al., 2022](#); [Bosticardo et al., 2022](#); [Kamagata et al., 2019](#); [Wei et al., 2018](#)). The tract qMRI estimates in all these studies are likely confounded by partial volume effects from crossing and kissing fibers which are present in 60-90% of white matter voxels ([Jeurissen et al., 2013](#)), which can potentially bias the resulting measures and could reduce the sensitivity to subtle differences. Incorporating tract-specific information from dual-encoded sequences would lead to more anatomically specific tract properties and more informative connectomes. However, the diffusion-relaxometry implementations described above often require time consuming acquisitions, advanced gradient performance, and/or complex processing routines, making them less amenable for use in patient populations.

Here we introduce an efficient dual-encoded magnetization transfer (MT) and DWI sequence to estimate the MT ratio (MTR) of individual white matter tracts using a global, whole brain optimization framework. MT is a contrast mechanism that is sensitive to the properties of bound macromolecular protons, such as bound pool fraction and exchange rate. In the brain, these bound macromolecules are largely found in cellular membranes including the lipid-rich myelin sheath (see review ([Sled, 2018](#))). Through simulations, the acquisition parameters of the dual-encoded sequence were optimized to co-encode this information in a clinically acceptable scan time of 7 mins at 2.6 mm isotropic voxel size at 3 Tesla. The optimal protocol was executed and repeated on 10 healthy subjects and analyzed using Convex Optimization Modeling for Microstructure Informed Tractography (COMMIT,([Daducci et al., 2015](#))) to map the distinct MTR values of fiber bundles in the tractogram. The tract-specific MTR values and their scan-rescan repeatability were compared to conventional MTR tractometry.

## Methods

### Sequence design:

The dual-encoded sequence is comprised of a spatially non-selective MT preparation module, inserted prior to the excitation of each slice in a 2D diffusion-weighted spin echo planar imaging (EPI) acquisition (Figure 1). A dual-polarity pulsed MT preparation module was used to maximize contrast while maintaining an acceptable radio-frequency (RF) power deposition ([Varma et al., 2018](#)). The following parameters can be controlled by the user at the console: the offset frequency and polarity (positive or alternating), pulse duration ($\tau$), inter-pulse time gap ($\Delta t$), number of pulses, and the flip angle of the MT pulse ($FA_{MT}$), as well as the duration of the spin echo excitation ($T_{exc}$) and refocusing pulses ($T_{ref}$). Previously implemented MT-weighted spin echo (SE)-EPI sequences used either a single Gaussian off-resonance pulse combined with diffusion weighting (Gupta et al., 2005) or used multiple pulses without diffusion weighting (Battiston et al., 2019).

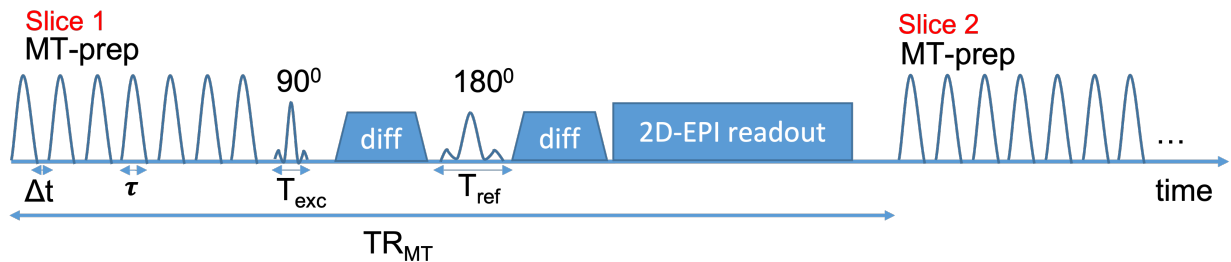

*Figure 1: Sequence diagram of co-encoded MT-diffusion: A pulsed MT module inserted prior to the diffusion preparation of the 2D-EPI acquisition of each slice. The duration ($\tau$), inter-pulse time gap ($\Delta t$), number of pulses, and the flip angle of the MT pulse ($FA_{MT}$) can be controlled as well as the duration of the excitation ($T_{exc}$) and refocusing pulses ($T_{ref}$)*

In contrast to the typical 3D spoiled gradient echo sequences used for MT-weighted experiments, diffusion acquisitions typically use SE-EPI with fat saturation and multi-band pulses (MB), which all contribute to SAR and thus limit the energy deposition that can be used for MT contrast. The MTR is given by the ratio of the images with and without MT saturation (MTR=1-$MT_{on}/MT_{off}$). This means that unwanted off-resonance contributions from the fat saturation and MB pulses will limit the available contrast since they are present in both $MT_{on}$ and $MT_{off}$ images. To maximize MT contrast, we chose to avoid MB acceleration at the cost of scan time. Furthermore, we investigated doing fat suppression by adjusting the ratio of the timing and amplitude of the excitation and refocusing pulses (Ivanov et al., 2010) rather than using a fat saturation pulse. These changes come at a cost of a slight increase in echo time (TE).

The number of slices and diffusion directions were chosen to provide full brain coverage and sufficient angular resolution for tractography, such that we can use a global optimization framework to estimate tract-specific MTR values (Daducci et al., 2015). We chose a relatively low b-value of 1500 s/mm² to maintain sufficient signal from the extra-axonal compartment, where MT contrast will arise from the interactions at the surface of the myelin sheath. The MT-weighted signal from myelin water has mostly decayed at the relatively long TE values (~60-70 ms) required for this diffusion weighting ([van Gelderen & Duyn, 2019](#)), whereas the MT contrast from the intra- and extra-axonal spaces will remain.

Simulations for sequence optimization:

To determine the acquisition parameters that maximize MT contrast efficiency, simulations of a 2-pool model, including a dipolar component, were carried out in MATLAB using recently presented optimization software ([Rowley C, 2022](#)) following a minimal approximation approach ([Portnoy & Stanisz, 2007](#)) (model assumptions: super Lorentzian lineshape for the bound pool, longitudinal relaxation time of the bound pool T1b = 1 s, transverse relaxation of the bound pool T2b = 1 µs, dipolar order relaxation time T1d = 3 ms, transverse relaxation of the free pool T2a = 60 ms, exchange rate between the pools R = 26 s$^{-1}$, bound pool fraction M0b = 0.1, observed relaxation time Raobs = 850 ms, Gaussian MT pulse shape).

The following parameter search space was simulated: MT offset frequency = 1-10 kHz; TR$_{MT}$ = 90-150 ms; number of pulses = 1-15; pulse duration = 1-12 ms. The following protocol parameters were kept fixed: Δt = 0.3 ms, resolution = 2.6 mm³, 62 slices, TE = 58 ms, b-value = 1500 s/mm², directions = 30. The FA$_{MT}$ was set to the maximum within the SAR constraints for the whole sequence (3.2 W/kg for the head). In addition, the total scan time was constrained to a maximum of 10 minutes, which includes the 2 acquisitions, with and without the MT preparation module, needed to compute the MTR. MTR represents the relative decrease in signal due to the saturation pulses, which can in part be due to the direct saturation of water and may lead to an erroneous attribution of changes to the macromolecular pool. To maximize our sensitivity to the macromolecular pool, MTR efficiency (MTR/scan time) was computed using simulations with the 2-pool model, where MTR is the difference between two cases where M0b = 0.1, and M0b = 0, with the latter representing the free-water pool only.

### In-vivo acquisitions

All MR images were acquired on a Siemens Prisma-Fit 3 Tesla scanner using a 32-channel head coil at the McConnell Brain Imaging Centre of the Montreal Neurological Institute. The project was reviewed and approved by the Research Ethics Board of McGill University.

The simulation results were verified with two in vivo datasets: first using the optimal protocol that was found to maximize MT contrast efficiency (offset frequency = 3 kHz; dual irradiation; $TR_{MT}$ = 90 ms; number of pulses = 7; pulse duration = 1 ms), and second with a higher $TR_{MT}$ = 110 ms while keeping the total SAR constant at 97% of the allowable limit. For all subsequent in-vivo acquisitions, the optimal MT preparation was used.

To verify the reduction of unwanted sources of SAR and off-resonance effects that arise from the MB excitation and refocusing pulses and fat saturation that are typically used in diffusion acquisitions, three different datasets were acquired. In the first, standard parameters were used (MB = 3, TE/TR = 55/3000 ms and fat saturation), in the second MB was removed resulting in a longer TR (TE/TR = 55/6400 ms, fat saturation), and in the third the fat saturation was replaced by an adjustment of the excitation and refocusing pulse duration ratios that minimizes refocusing of the fat signal. The pulse lengths were computed according to Eqn 8 in Ivanov et al. (Ivanov et al., 2010) for a field strength of 3 T and a slice thickness of 2.6 mm (TE/TR = 58/5900 ms, $T_{exc}$ = 3.328 ms, $T_{ref}$ = 9.472 ms). In all cases, the MT preparation was kept the same (optimal based on simulations), the TE and TR were set to the minimum, and the $FA_{MT}$ was increased until the total SAR reached 97% of the allowable limit. All other parameters were kept constant (63 slices, GRAPPA = 2, resolution = 2.6mm isotropic, PF = 6/8, BW = 1500 Hz/px, b-value = 1500 s/mm$^2$).

The optimal parameter combination was used to scan 10 healthy subjects (4 women, aged 30 +/- 10 years) over 2 separate sessions in order to assess scan-rescan repeatability. The acquisitions included the optimal MT-diffusion ($MT_{on}$ and $MT_{off}$), a reverse phase-encoded b=0 scan for distortion correction, and a T1-weighted MPRAGE for registration and brain tissue segmentation (1 mm isotropic, GRAPPA = 2, TE/TI/TR = 2.98/900/2300 ms, FA = 8°).

## Image pre-processing

Diffusion weighted images (DWIs) (both with and without MT saturation) were denoised (Veraart et al., 2016) and pre-processed to account for subject movement, susceptibility and eddy current induced distortions with a combination of MRtrix3 (Tournier et al., 2019) and FSL (Andersson et al., 2003). Image non-uniformity correction was performed with ANTs (Tustison et al., 2010), using the bias field estimated from the $MT_{off}$ data applied to both the $MT_{off}$ and $MT_{on}$ data; the same correction was used for both acquisitions. Subsequently, all DWIs were up-sampled to 1 mm isotropic voxels, registered to the T1-weighted MPRAGE image and aligned to the Desikan-Killiany segmentation atlas using FreeSurfer v7 (Desikan et al., 2006). The brainstem was further segmented into substructures (Iglesias et al., 2015). The computed transformations were inverted and applied to the T1-weighted image and the atlas such that all further analysis was carried out in the subject's native diffusion space. Anatomically constrained probabilistic tractography was performed on the $MT_{off}$ data only using the iFOD2 algorithm (J. Tournier, 2010) with 3 million streamlines, as implemented in MRtrix3. Streamlines not connecting nodes of the atlas were discarded.

## Tract-specific MTR analysis pipeline

COMMIT was used to estimate tract-specific MTR values as follows. COMMIT estimates a chosen parameter describing a microstructural property of a streamline from tractography with the assumption that this parameter is constant along the streamline's trajectory. This parameter is called the streamline weight, $x$. In the current implementation, the streamline weight is the signal per unit length of the part of the fiber bundle represented by the streamline. Ignoring sources of signal variation such as relaxation and macromolecular content, this can be interpreted as a volume per unit length, i.e., the cross-sectional area (Smith et al., 2022). Here, the compartments of the fiber bundle attributed to the streamline consists of the combined intra- and extra-axonal spaces of the fiber. The diffusion response function of this combined space can be represented by an anisotropic tensor or "zeppelin". The other space in the voxel is represented by a "ball" modeling free water (Panagiotaki et al., 2012) and can vary from voxel to voxel. It is therefore assumed that the signal contribution from the intra-axonal space and its immediate surrounding extra-axonal space is constant along each streamline. Furthermore, all streamlines throughout the brain are assumed to have the same diffusivity parameters (diffusivity parallel to the streamline direction $D_\parallel$ = 1.7E-3 mm²/s; diffusivity perpendicular to the streamline direction $D_\perp$ = 0.6E-3 mm²/s) and the ball compartment (isotropic diffusivity= 3E-3 mm²/s to capture contributions from free water). COMMIT was applied separately to the $MT_{on}$ and $MT_{off}$ datasets, using the tractogram computed on the $MT_{off}$ dataset. In both cases, the signal was normalized to the b=0 s/mm² of the $MT_{off}$ dataset. The drop in signal due to MT-weighting will lead to a proportional change in streamline weights between the two fits. The following voxel-wise equations describe the fitting process of the $MT_{on}$ and $MT_{off}$ datasets:

$$\frac{S(q, MT_{off})}{S(b=0, MT_{off})} = \sum_i x_i^{zeppelin,MT_{off}} R_i^{zeppelin}(q) + x^{ball,MT_{off}}$$

$$\frac{S(q, MT_{on})}{S(b=0, MT_{off})} = \sum_i x_i^{zeppelin,MT_{on}} R_i^{zeppelin}(q) + x^{ball,MT_{on}}$$

where S(**q**) is the signal at each q-space location, $R^{zeppelin}$ represents the response function of the zeppelin compartment, rotated to align to the fiber orientation, and is scaled by the length of the streamline intersecting the voxel. Finally, the $x^{zeppelin}$ represent the contribution of each streamline (*i*) and $x^{ball}$ the contribution of free water in the voxel. Streamlines whose weight was greater than zero in both $MT_{on}$ and $MT_{off}$ were grouped into bundles according to the pair of grey matter regions they connect. The tract-specific $MT_{on}$ and $MT_{off}$ were then calculated as the sum of all streamline volumes in the bundle. The volumetric contribution of streamline *j* to the bundle is given by the product of its weight $x_j$ times its length $L_j$. Finally, the tract-specific MTR is computed by combining the individual connectomes with the standard equation ([S. Schiavi et al., 2020](#)):

$$MTR_{bundle} = 1 - MTon_{bundle}/MToff_{bundle} = 1 - \frac{\sum_j^N L_j \cdot x_j^{zeppelin,MT_{on}}}{\sum_j^N L_j \cdot x_j^{zeppelin,MT_{off}}}$$

A selection of large white matter tracts that connect nodes of the Desikan-Killiany atlas were extracted from the final connectome. The selected tracts are listed in Table 1 along with the nodes they connect.

| Name | Description | Name of brain region and (node number) |
| --- | --- | --- |
| fWM | pathways connecting the bilateral frontal white matter, passing through the genu of the corpus callosum | Left (27) to right (76) superior frontal cortex |
| SLF | Superior longitudinal fasciculus | Left (27) superior frontal to left (28) superior parietal cortex<br>Right (76) superior frontal to right (77) superior parietal cortex |
| Pons | pontine fibers | Left (35) and right (84) cerebellar cortex |
| CST | pathways of the corticospinal tract that connect the brainstem and motor cortex | Brainstem to left (23) and right (72) precentral gyrus |
| PrCG-Thal | projection pathway between the precentral gyrus and thalamus | Left precentral gyrus (23) to left thalamus (36). Right precentral gyrus (72) to right thalamus (43). |
| Splen | pathways connecting the bilateral occipital white matter, passing through the splenium of the corpus callosum | Left lateral occipital cortex (10) to right lateral occipital cortex (59) |

*Table 1: Selected bundle names, description, and nodes of the Desikan-Killiany parcellation that they connect*

The pipeline is illustrated in Figure 2.

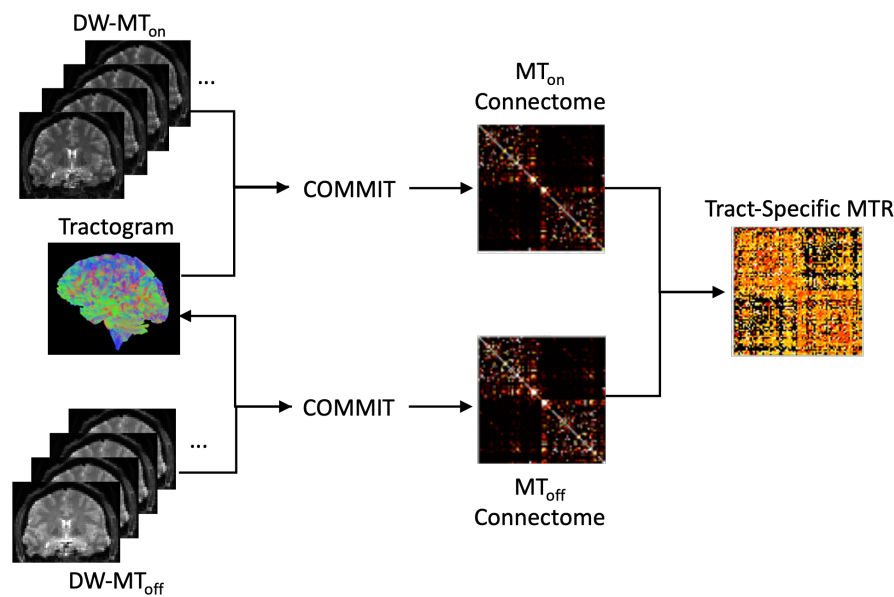

*Figure 2: Processing pipeline: The $MT_{off}$ data is used to generate the tractogram, which is then used with COMMIT to produce the $MT_{on}$ and $MT_{off}$ connectomes. These connectomes are then combined to get tract-specific MTR values*

As previously done in Schiavi et al 2022 (Schiavi et al., 2022), the proposed method was compared to what we will refer to henceforth as conventional tractometry, whereby a separate MTR map was sampled along each streamline, taking the median along its length, and averaged across streamlines within a bundle. The MTR maps used here were calculated as the average across all directions of the diffusion-weighted datasets, omitting the b=0 image ($MTR_{dw}$). For both methods, subject-wise bundle MTR values were compared with a t-test, to quantify whether there were consistent differences between bundles. The full pipeline is available on github (https://github.com/TardifLab/mt-diff) and data is available via request through a formal data sharing agreement and approval from the local ethics committees.

## Results

### Sequence optimization

Based on the simulation results (Figure 3), our optimal protocol was: offset frequency = 3 kHz; dual irradiation; $TR_{MT}$ = 90 ms (minimum); number of MT pulses = 7; pulse duration = 1 ms. These results point to $TR_{MT}$ being the parameter with the most significant impact on MT contrast efficiency. This is seen in Figure 3A and B, where for a fixed $TR_{MT}$, several different combinations of offset frequencies and number of pulses lead to a similar efficiency. The simulation results were validated by acquiring 2 datasets, first using the optimal protocol with $TR_{MT} = 90$ ms (red circle in Figure 3), and second at a higher $TR_{MT} = 110$ ms (green circle) while keeping the total SAR constant at 97% of the allowable limit. This resulted in a $FA_{MT}$ of 596° and 656° respectively.

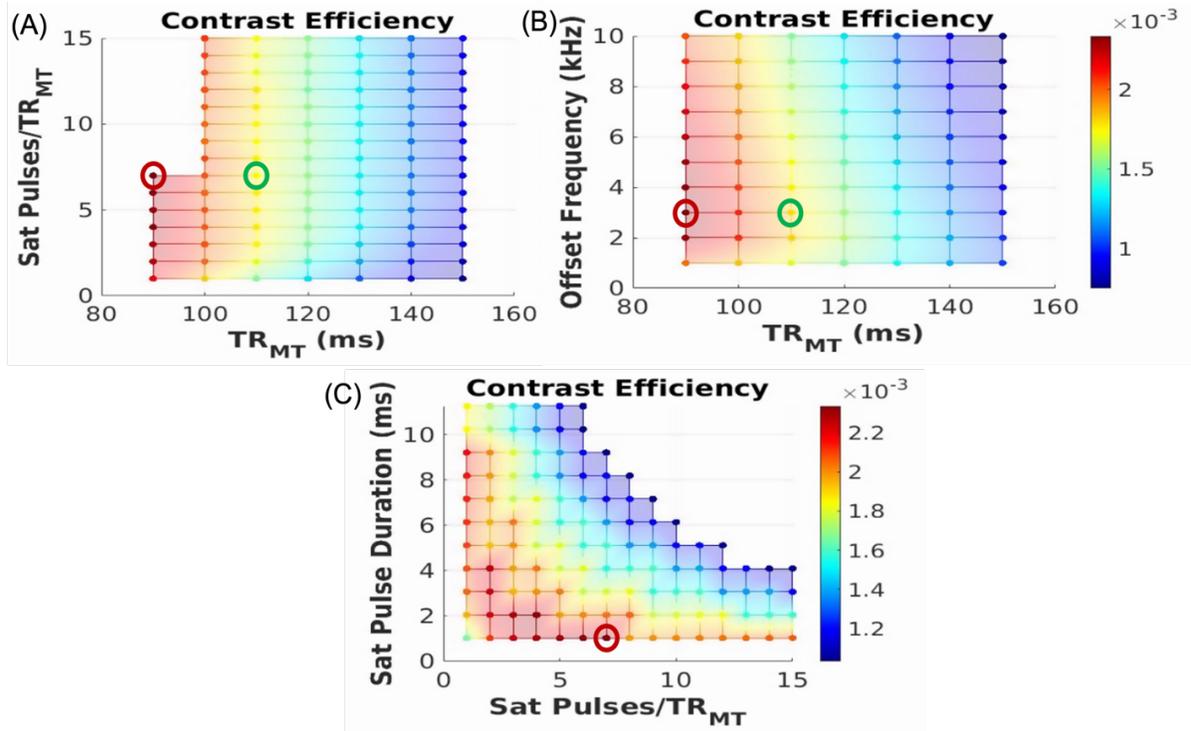

*Figure 3: Simulated results of contrast efficiency: (A) Number of saturation pulses vs $TR_{MT}$ (B) Offset frequency vs $TR_{MT}$ and (C) Saturation pulse durations vs number of saturation pulses. Red circle highlights the optimal protocol and the green, sub-optimal*

Figure 4 illustrates the contrast efficiency (MTR/time) calculated using the b=0 images with and without MT-weighting, as well as for the average of 6 of the 30 diffusion orientations (Diff$_{AVG}$). This agrees with the simulation results, where minimizing the $TR_{MT}$ increases the contrast efficiency.
12

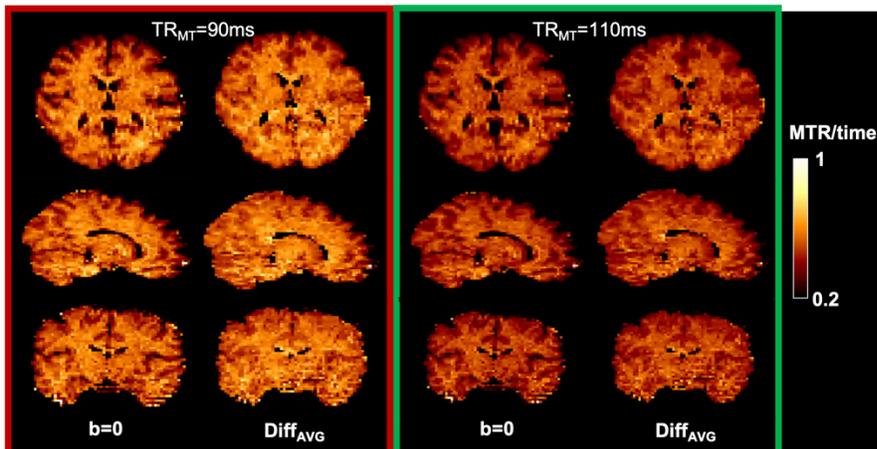

*Figure 4: Contrast efficiency (MTR/time) of optimal (TR$_{MT}$ = 90ms, red outline) and a sub-optimal (TR$_{MT}$ = 110ms, green outline) protocols of the b=0 and of the average over 6 diffusion directions (Diff$_{AVG}$)*

As shown in an example diffusion-weighted image (Figure 5A), the scalp fat signal was successfully suppressed with a T$_{exc}$ = 3.328 ms and T$_{ref}$ = 9.472 ms for a slice thickness of 2.6 mm, at the cost of a slight increase in TE (+3 ms) compared to the standard protocol using fat saturation. The TR is effectively reduced with the fat shifting technique, since the 5.1 ms fat saturation pulse prior to each slice is removed (6400 ms vs 5900 ms). Figure 5B illustrates the gain in MTR contrast (68%) that can be achieved by removing MB and the standard fat saturation, whereby the FA$_{MT}$ can be increased while remaining within 97% of the allowable SAR limit.

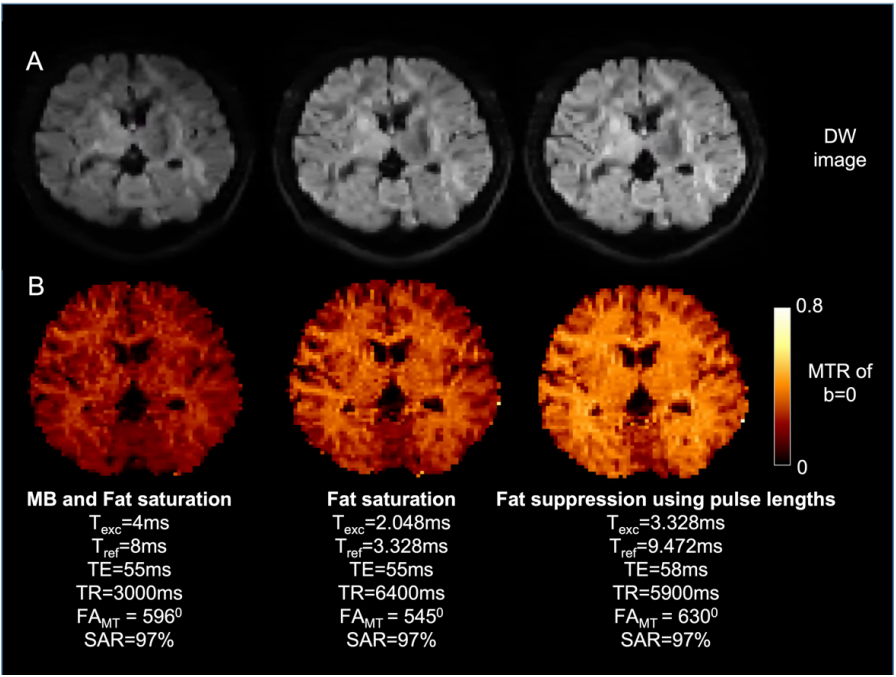

*Figure 5: Reducing sources of unwanted SAR and MT by avoiding MB and replacing standard fat saturation by using a ratio of pulse lengths for fat suppression. (A) diffusion-weighted image (B) the MTR of the b=0 images*

### In vivo comparison of tract-specific and tractometry MTR

The results for tract-specific and tractometry MTR values across all 10 subjects are shown in Figure 6 for the selected white matter tracts listed in Table 1. As shown in (A), there are significant within subject differences between the methods for the fWM, Pons and Splen (shown with an *, p<0.01). In (B) and (C), the group-wise MTR values are shown for both methods, as well as the corresponding scan-rescan repeatability in (D) and (E). The mean scan-rescan percent difference of tract-specific MTR in the selected tracts is slightly higher than that for tractometry (~3% vs ~1%) and as expected, the bilateral differences (left (L) and right (R)) are not significant (n.s.) for both methods. However, the tract-specific MTR exhibits a higher dynamic range and a significant difference between the pons and CST which is not present for the standard tractometry method. For both methods, there is a significant difference between the CST and PrCG-Thal connections.

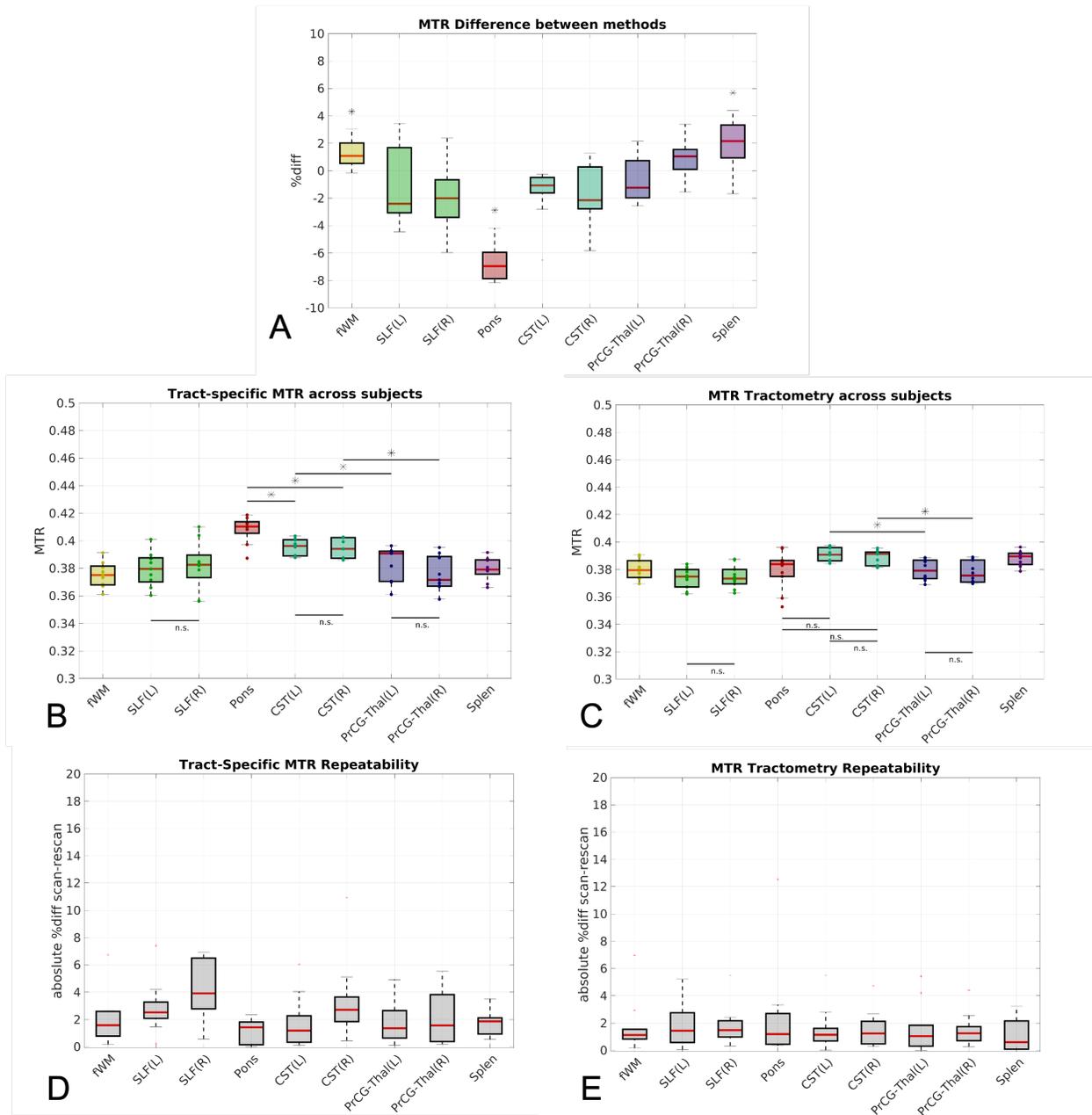

*Figure 6: (A) Percent difference between tract-specific and tractometry MTR for selected bundles (B) Average tract-specific MTR (C) Average tractometry MTR and (D,E) corresponding scan-rescan repeatability across 10 subjects (* denotes significance at p<0.01; n.s. not significant)*

An example of these intersecting tracts is shown in Figure 7, where the difference in tract-to-tract variability between the two methods is evidenced. The MTR results are overlayed on the $MTR_{dw}$ map for a single subject, echoing the trend in tract-specific MTR values of Figure 6B (Figure 7A: $MTR_{pons} > MTR_{CST}$; Figure 7B: $MTR_{CST} > MTR_{PrCG-Thal}$). The zoomed-in boxes in Figure 7 highlight regions where a single tract population is dominating the voxel, such that the value in the underlying MTR$_{dw}$ map gives a good indication of the expected MTR along its length. For example, the yellow open arrows in panel A point to voxels where the pons is the dominating fiber population and the red arrows point to voxels where the CST is the dominant population, such that the corresponding scalar MTR$_{dw}$ is indicative of the non-partial-volumed MTR of the tract. This map shows higher values in the pons than in the CST (blue color scale). The tract-specific MTR result is in better agreement with these values when compared to the tractometry results, which show less difference in MTR values across bundles. In both the MT$_{on}$ and MT$_{off}$ datasets, the contribution of the ball compartment in white matter was negligible, implying that the effects of CSF contamination on the overall fitting are likely minimal.

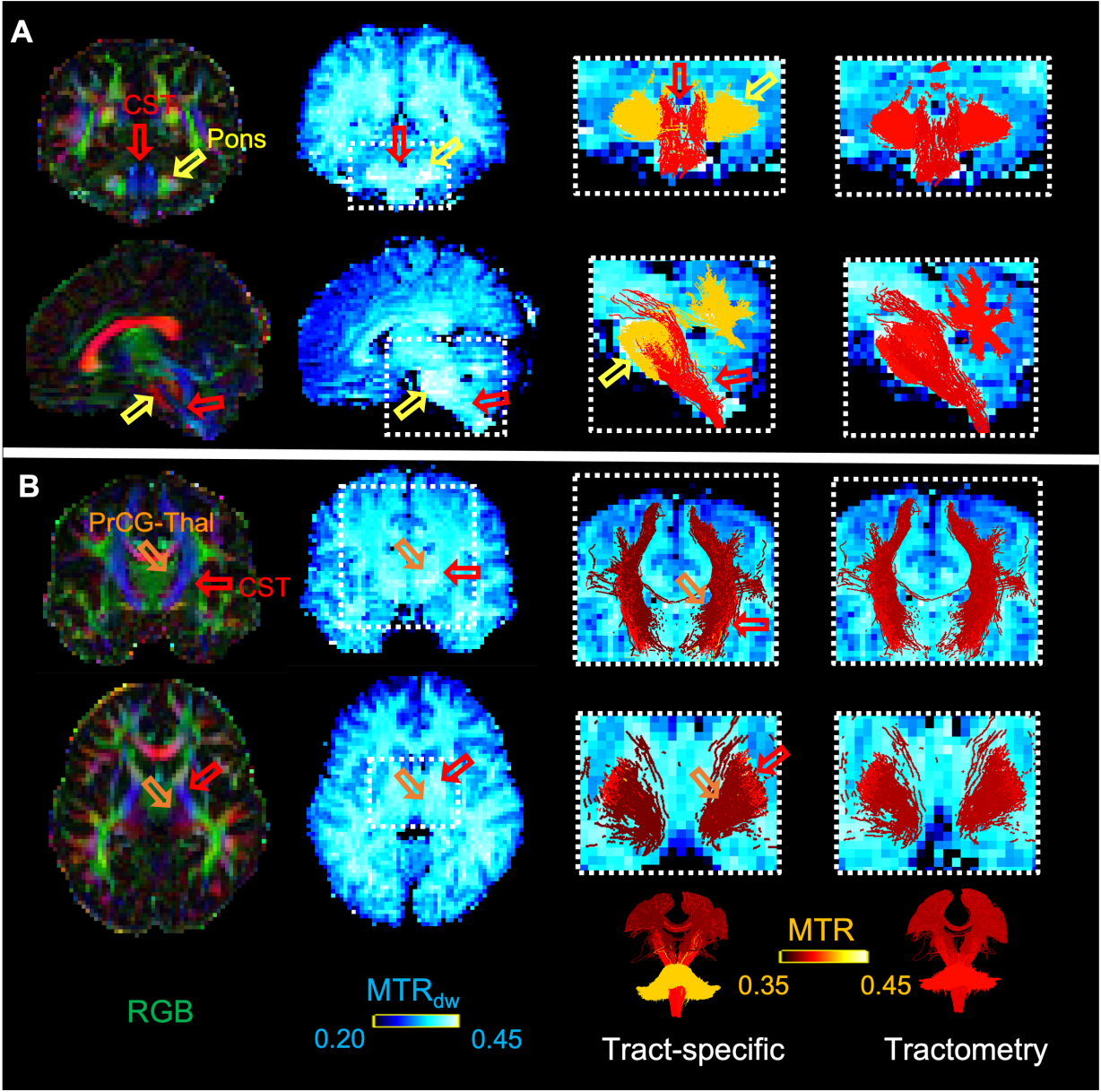

*Figure 7: Example of tract-specific and tractometry bundle MTR results for one subject.*

*Open arrows highlight the regions where a single fiber population is dominating the voxel (for example in (A): the pons (yellow) and CST (red) and (B): the PrCG-Thal connection (orange)), which gives an indication of the non-partial-volumed MTR$_{dw}$ value along each tract (blue color scale). The overlayed MTR results (hot color scale) for the tract-specific method show better agreement with the underlying scalar map and higher contrast between tracts compared to conventional tractometry.*

## Discussion:

The goal of this work was to design and optimize an efficient dual-encoded sequence, which can be combined with a global optimization microstructure informed tractography framework to estimate tract-specific MTR values. One of the primary motivations is to be able to provide tract-specific myelin indices, which should in turn lead to more anatomically and microstructurally specific myelin-weighted structural connectomes.

In terms of the sequence design, our simulation results agree with previous work from Varma et al ([Varma et al., 2018](#)) where the use of dual irradiation and rapidly switching between polarities, in this case short 1ms pulses with alternating polarity, help maximize MTR contrast ([Lee et al., 2011](#); [Varma et al., 2017](#)). In addition, we showed that by not using MB pulses and fat saturation, more power can be used for MT contrast, at the cost of an increase in scanning time. Nevertheless, the optimized protocol with both $MT_{on}$ and $MT_{off}$, can be acquired in under 7 minutes.

In line with previous literature ([Garcia et al., 2011](#); [Mehta et al., 1995](#)), the range of MTR values in major white matter fiber tracts in healthy young adults is relatively small. This and the low number of subjects (10) might explain the limited number of significant differences between tracts at the subject level for both the tract-specific and tractometry methods. However, in the case of the tract-specific values, the dynamic range is larger, and the group trends are corroborated by the $MTR_{dw}$ map, particularly when referring to regions where a particular tract is the dominating population in the voxel. Conversely, the tractometry maps are relatively flat, likely due to the extensive partial volume effect occurring over the length of the tracts. The increase in dynamic range has also been seen in previous work using both voxel-wise and global approaches compared to more standard tractometry ([De Santis et al., 2016](#); [Leppert et al., 2021](#); [Schiavi et al., 2022](#)). The scan-rescan repeatability is lower in the tract-specific approach compared to the tractometry, which is expected given the increased sensitivity of per-streamline fitting compared to the blurring that sampling and averaging along an underlying scalar map entail. Sources of error that particularly affect the global optimization technique are false positives and false negatives in tractography and the sensitivity to $B1^+$ non-uniformity. Future work will focus on exploring model-based corrections for $B1^+$ non-uniformity ([Rowley et al., 2021](#)) and the reduction of $T_1$ bias in the MTR maps ([Helms et al., 2008](#)).

Previous work on estimating tract-specific measures includes co-encoded voxel-wise fitting of myelin-sensitive $T_1$ (De Santis et al., 2016), co-encoded global tract-based analysis of axon caliber (Barakovic, Girard, et al., 2021) and intra-axonal $T_2$ (Barakovic, Tax, et al., 2021) as well as global tract-based analysis to estimate bundle-specific myelin content using separate acquisitions (MTsat and myelin water fraction) for myelin contrast (Simona Schiavi et al., 2020). The current implementation aims to draw advantages from these previous methods while attempting to simplify both the acquisition and analysis by using fewer measurements and a simple ratio of contrasts with global tractograms, while still providing co-encoded information. One advantage of the global optimization technique over voxel-wise approaches is its ability to dissociate bundles that are parallel at the voxel level for some extent of their trajectory and then diverge, becoming geometrically distinct globally. The co-encoding approach presented here is expected to provide additional potential for bundle dissociation, particularly when there are large expanses of voxels with multiple bundles, which is the case for most of the voxels in the brain (Jeurissen et al., 2013). If partial volume effects were not an issue, tractometry would theoretically have given the same results. COMMIT can dissociate parallel tracts because of its global cost function, which attempts to fit the signal in all voxels simultaneously, combined with the assumption that the microstructure of a streamline is constant along its length.

The choice of the zeppelin & ball model was made primarily because the MT effect is expected to be significant in the extra-axonal space, and we want to capture all of this. Unlike other implementations of COMMIT (Daducci et al., 2015) that keep the intra-axonal compartment signal constant along all streamlines and the extra-axonal compartment varies per fixel (fiber element in a voxel), we chose to keep the intra- plus proximal extra-axonal compartments constant. This implies that changes in fiber density within a single streamline are minimal or have a minimal effect on the diffusion signal.

Although the MTR contrast reported here will correlate with myelin density to some extent, it will be modulated by the surface to volume ratio, sheath thickness, and exchange between compartments, and thus be sensitive to fiber size, the presence of other bound proton populations, orientation and packing geometry as well. In fact, the source of the MTR contrast is different than traditional MT-GRE experiments, whose short TE is geared towards weighting interactions at both the inner and outer surface of the myelin as well as within the sheath in order to estimate myelin content. With a longer TE and diffusion weighting, the MT-weighted signal is reduced overall, but to a different extent for each compartment. The echo time will modulate the contribution of each compartment based on their respective $T_2$, such that the myelin water ($T_2$ = 10-40ms, (MacKay et al., 1994)) signals contribute the least and the extra-axonal ($T_2$ = 30-50ms) and intra-axonal ($T_2$ = 80-120ms) (Veraart et al., 2018) signals contribute the most. Diffusion weighting will have a greater attenuation on the extra-axonal water than the intra-axonal compartment. In addition, the MT-diffusion experiment is inherently less affected by unwanted direct saturation effects, due to the attenuation of free water by diffusion and the fact that it is modeled as a separate compartment (ball). Therefore, through the modulation of TE and b-value, the contributions of the different compartments to the tract-specific MTR measurements can be explored. Centric k-space encoding techniques and powerful gradients can be used to shorten the TE and increase the achievable b-value to gain a more detailed understanding of the origin of the MT- and diffusion-weighted signal.

Ultimately, many of the methods discussed above provide complementary tract-specific measures of microstructure that could be combined to understand the role of tract microstructure on brain connectivity and function. Future work will include comparing their performance across different modalities and applications.

## Conclusion

This work presents a novel dual-encoded MT and diffusion sequence, for which parameters have been optimized for MT contrast efficiency. The resulting 2.6 mm whole brain protocol can be acquired in under 7 minutes and is an important step towards providing tract-specific myelin indices that minimize biases due to partial volume effects with neighboring tracts. In turn, this will provide more statistical power and insight when the microstructure of specific tracts is altered, for example through disease, ageing, function, or treatment. Finally, the potential to provide more anatomically-specific connectomes could have a significant impact on brain network analysis.

## Acknowledgments:

This work was supported by the following funding sources: Quebec Bio-Imaging Network (QBIN), Fonds de Recherche du Québec – Santé (FRQS), CFREF Healthy Brains for Healthy Lives (HBHL), NSERC (RGPIN-03880, RGPIN-2018-05176), Brain Canada. The authors have no conflicts of interest to declare.

## References:

Andersson, J. L., Skare, S., & Ashburner, J. (2003). How to correct susceptibility distortions in spin-echo echo-planar images: application to diffusion tensor imaging. *NeuroImage*, *20*(2), 870-888. https://doi.org/10.1016/S1053-8119(03)00336-7

Barakovic, M., Girard, G., Schiavi, S., Romascano, D., Descoteaux, M., Granziera, C., Jones, D. K., Innocenti, G. M., Thiran, J.-P., & Daducci, A. (2021). Bundle-Specific Axon Diameter Index as a New Contrast to Differentiate White Matter Tracts. *Frontiers in Neuroscience*, *15*. https://doi.org/10.3389/fnins.2021.646034

Barakovic, M., Tax, C. M. W., Rudrapatna, U., Chamberland, M., Rafael-Patino, J., Granziera, C., Thiran, J. P., Daducci, A., Canales-Rodríguez, E. J., & Jones, D. K. (2021). Resolving bundle-specific intra-axonal T2 values within a voxel using diffusion-relaxation tract-based estimation. *NeuroImage*, *227*. https://doi.org/10.1016/j.neuroimage.2020.117617

Battiston, M., Schneider, T., Grussu, F., Yiannakas, M. C., Prados, F., De Angelis, F., Gandini Wheeler-Kingshott, C. A. M., & Samson, R. S. (2019). Fast bound pool fraction mapping via steady-state magnetization transfer saturation using single-shot EPI. *Magnetic Resonance in Medicine*, *82*(3). https://doi.org/10.1002/mrm.27792

Bells, S., Cercignani, M., Deoni, S., Assaf, Y., Pasternak, O., Evans, J. C., Leemans, A., & Jones, D. (2011). Tractometry comprehensive multi-modal quantitative assessment of white matter along specific tracts. Annual Meeting of the International Society for Magnetic Resonance in Medicine, Montreal, Canada.

Benjamini, D., & Basser, P. J. (2016). Use of marginal distributions constrained optimization (MADCO) for accelerated 2D MRI relaxometry and diffusometry. *J Magn Reson*, *271*, 40-45. https://doi.org/10.1016/j.jmr.2016.08.004

Benjamini, D., & Basser, P. J. (2020). Multidimensional correlation MRI. *NMR Biomed*, *33*(12), e4226. https://doi.org/10.1002/nbm.4226

Boshkovski, T., Cohen-Adad, J., Misic, B., Arnulf, I., Corvol, J. C., Vidailhet, M., Lehericy, S., Stikov, N., & Mancini, M. (2022). The Myelin-Weighted Connectome in Parkinson's Disease. *Mov Disord*, *37*(4), 724-733. https://doi.org/10.1002/mds.28891

Bosticardo, S., Schiavi, S., Schaedelin, S., Lu, P. J., Barakovic, M., Weigel, M., Kappos, L., Kuhle, J., Daducci, A., & Granziera, C. (2022). Microstructure-Weighted Connectomics in Multiple Sclerosis. *Brain Connect*, *12*(1), 6-17. https://doi.org/10.1089/brain.2021.0047

Callaghan, M. F., Freund, P., Draganski, B., Anderson, E., Cappelletti, M., Chowdhury, R., Diedrichsen, J., Fitzgerald, T. H., Smittenaar, P., Helms, G., Lutti, A., & Weiskopf, N. (2014). Widespread age-related differences in the human brain microstructure revealed by quantitative magnetic resonance imaging. *Neurobiol Aging*, *35*(8), 1862-1872. https://doi.org/10.1016/j.neurobiolaging.2014.02.008

Cercignani, M., & Bouyagoub, S. (2018). Brain microstructure by multi-modal MRI: Is the whole greater than the sum of its parts? *NeuroImage*, *182*, 117-127. https://doi.org/10.1016/j.neuroimage.2017.10.052

Correia, S., Lee, S. Y., Voorn, T., Tate, D. F., Paul, R. H., Zhang, S., Salloway, S. P., Malloy, P. F., & Laidlaw, D. H. (2008). Quantitative tractography metrics of white matter integrity in diffusion-tensor MRI. *NeuroImage*, *42*(2), 568-581. https://doi.org/10.1016/j.neuroimage.2008.05.022

Daducci, A., Dal Palù, A., Lemkaddem, A., & Thiran, J. P. (2015). COMMIT: Convex optimization modeling for microstructure informed tractography. *IEEE Transactions on Medical Imaging*, *34*(1). https://doi.org/10.1109/TMI.2014.2352414

Dayan, M., Monohan, E., Pandya, S., Kuceyeski, A., Nguyen, T. D., Raj, A., & Gauthier, S. A. (2016). Profilometry: A new statistical framework for the characterization of white matter pathways, with application to multiple sclerosis. *Hum Brain Mapp*, *37*(3), 989-1004. https://doi.org/10.1002/hbm.23082

de Almeida Martins, J. P., Tax, C. M. W., Reymbaut, A., Szczepankiewicz, F., Chamberland, M., Jones, D. K., & Topgaard, D. (2021). Computing and visualising intra-voxel orientation-specific relaxation-diffusion features in the human brain. *Hum Brain Mapp*, *42*(2), 310-328. https://doi.org/10.1002/hbm.25224

De Santis, S., Barazany, D., Jones, D. K., & Assaf, Y. (2016). Resolving relaxometry and diffusion properties within the same voxel in the presence of crossing fibres by combining inversion recovery and diffusion-weighted acquisitions. *Magnetic Resonance in Medicine*, *75*(1), 372-380. http://view.ncbi.nlm.nih.gov/pubmed/25735538

Desikan, R. S., Segonne, F., Fischl, B., Quinn, B. T., Dickerson, B. C., Blacker, D., Buckner, R. L., Dale, A. M., Maguire, R. P., Hyman, B. T., Albert, M. S., & Killiany, R. J. (2006). An automated labeling system for subdividing the human cerebral cortex on MRI scans into gyral based regions of interest. *NeuroImage*, *31*(3), 968-980. https://doi.org/10.1016/j.neuroimage.2006.01.021

Does, M. D. (2018). Inferring brain tissue composition and microstructure via MR relaxometry. *NeuroImage*. http://view.ncbi.nlm.nih.gov/pubmed/29305163

Garcia, M., Gloor, M., Bieri, O., Wetzel, S. G., Radue, E. W., & Scheffler, K. (2011). MTR variations in normal adult brain structures using balanced steady-state free precession. *Neuroradiology*, *53*(3), 159-167. https://doi.org/10.1007/s00234-010-0714-5

Gong, T., Tong, Q., He, H., Sun, Y., Zhong, J., & Zhang, H. (2020). MTE-NODDI: Multi-TE NODDI for disentangling non-T2-weighted signal fractions from compartment-specific T2 relaxation times. *NeuroImage*, *217*, 116906. https://doi.org/10.1016/j.neuroimage.2020.116906

Gupta, R. K., Rao, A. M., Mishra, A. M., Chawla, S., Sekar, D. R., & Venkatesan, R. (2005). Diffusion-weighted EPI with magnetization transfer contrast. *Magnetic Resonance Imaging*, *23*(1). https://doi.org/10.1016/j.mri.2004.11.005

Helms, G., Dathe, H., Kallenberg, K., & Dechent, P. (2008). High-resolution maps of magnetization transfer with inherent correction for RF inhomogeneity and T1 relaxation obtained from 3D FLASH MRI. *Magnetic Resonance in Medicine*, *60*(6). https://doi.org/10.1002/mrm.21732

Iglesias, J. E., Van Leemput, K., Bhatt, P., Casillas, C., Dutt, S., Schuff, N., Truran-Sacrey, D., Boxer, A., Fischl, B., & Alzheimer's Disease Neuroimaging, I. (2015). Bayesian segmentation of brainstem structures in MRI. *NeuroImage*, *113*, 184-195. https://doi.org/10.1016/j.neuroimage.2015.02.065

Ivanov, D., Schäfer, A., Streicher, M. N., Heidemann, R. M., Trampel, R., & Turner, R. (2010). A simple low-SAR technique for chemical-shift selection with high-field spin-echo imaging. *Magnetic Resonance in Medicine*, *64*(2). https://doi.org/10.1002/mrm.22518

J. Tournier, F. C., A. Connelly. (2010). Improved probabilistic streamlines tractography by 2nd order integration over fibre orientation distributions. Proceedings of the International Society for Magnetic Resonance in Medicine,

Jeurissen, B., Leemans, A., Tournier, J. D., Jones, D. K., & Sijbers, J. (2013). Investigating the prevalence of complex fiber configurations in white matter tissue with diffusion magnetic resonance imaging. *Hum Brain Mapp*, *34*(11), 2747-2766. https://doi.org/10.1002/hbm.22099

Kamagata, K., Zalesky, A., Yokoyama, K., Andica, C., Hagiwara, A., Shimoji, K., Kumamaru, K. K., Takemura, M. Y., Hoshino, Y., Kamiya, K., Hori, M., Pantelis, C., Hattori, N., & Aoki, S. (2019). MR g-ratio-weighted connectome analysis in patients with multiple sclerosis. *Sci Rep*, *9*(1), 13522. https://doi.org/10.1038/s41598-019-50025-2

Kim, D., Doyle, E. K., Wisnowski, J. L., Kim, J. H., & Haldar, J. P. (2017). Diffusion-relaxation correlation spectroscopic imaging: A multidimensional approach for probing microstructure. *Magn Reson Med*, *78*(6), 2236-2249. https://doi.org/10.1002/mrm.26629

Kolind, S. H., Laule, C., Vavasour, I. M., Li, D. K., Traboulsee, A. L., Madler, B., Moore, G. R., & Mackay, A. L. (2008). Complementary information from multi-exponential T2 relaxation and diffusion tensor imaging reveals differences between multiple sclerosis lesions. *NeuroImage*, *40*(1), 77-85. https://doi.org/10.1016/j.neuroimage.2007.11.033

Lampinen, B., Szczepankiewicz, F., Martensson, J., van Westen, D., Hansson, O., Westin, C. F., & Nilsson, M. (2020). Towards unconstrained compartment modeling in white matter using diffusion-relaxation MRI with tensor-valued diffusion encoding. *Magn Reson Med*, *84*(3), 1605-1623. https://doi.org/10.1002/mrm.28216

Lazari, A., & Lipp, I. (2021). Can MRI measure myelin? Systematic review, qualitative assessment, and meta-analysis of studies validating microstructural imaging with myelin histology. *NeuroImage*, *230*, 117744. https://doi.org/10.1016/j.neuroimage.2021.117744

Lee, J. S., Khitrin, A. K., Regatte, R. R., & Jerschow, A. (2011). Uniform saturation of a strongly coupled spin system by two-frequency irradiation. *J Chem Phys*, *134*(23), 234504. https://doi.org/10.1063/1.3600758

Leppert, I. R., Andrews, D. A., Campbell, J. S. W., Park, D. J., Pike, G. B., Polimeni, J. R., & Tardif, C. L. (2021). Efficient whole-brain tract-specific T1 mapping at 3T with slice-shuffled inversion-recovery diffusion-weighted imaging. *Magnetic Resonance in Medicine*, *86*(2). https://doi.org/10.1002/mrm.28734

Li, Y., Yan, S., Zhang, G., Shen, N., Wu, D., Lu, J., Zhou, Y., Liu, Y., Zhu, H., Li, L., Zhang, S., & Zhu, W. (2022). Tractometry-Based Estimation of Corticospinal Tract Injury to Assess Initial Impairment and Predict Functional Outcomes in Ischemic Stroke Patients. *J Magn Reson Imaging*, *55*(4), 1171-1180. https://doi.org/10.1002/jmri.27911

MacKay, A., Whittall, K., Adler, J., Li, D., Paty, D., & Graeb, D. (1994). In vivo visualization of myelin water in brain by magnetic resonance. *Magn Reson Med*, *31*(6), 673-677. https://doi.org/10.1002/mrm.1910310614

Mancini, M., Karakuzu, A., Cohen-Adad, J., Cercignani, M., Nichols, T. E., & Stikov, N. (2020). An interactive meta-analysis of MRI biomarkers of myelin. *Elife*, *9*. https://doi.org/10.7554/eLife.61523

Mangeat, G., Govindarajan, S. T., Mainero, C., & Cohen-Adad, J. (2015). Multivariate combination of magnetization transfer, T2* and B0 orientation to study the myelo-architecture of the in vivo human cortex. *NeuroImage*, *119*, 89-102. https://doi.org/10.1016/j.neuroimage.2015.06.033

Mehta, R. C., Pike, G. B., & Enzmann, D. R. (1995). Magnetization transfer MR of the normal adult brain. *AJNR Am J Neuroradiol*, *16*(10), 2085-2091. https://www.ncbi.nlm.nih.gov/pubmed/8585499

Panagiotaki, E., Schneider, T., Siow, B., Hall, M. G., Lythgoe, M. F., & Alexander, D. C. (2012). Compartment models of the diffusion MR signal in brain white matter: a taxonomy and comparison. *NeuroImage*, *59*(3), 2241-2254. https://doi.org/10.1016/j.neuroimage.2011.09.081

Portnoy, S., & Stanisz, G. J. (2007). Modeling pulsed magnetization transfer. *Magn Reson Med*, *58*(1), 144-155. https://doi.org/10.1002/mrm.21244

Reich, D. S., Zackowski, K. M., Gordon-Lipkin, E. M., Smith, S. A., Chodkowski, B. A., Cutter, G. R., & Calabresi, P. A. (2008). Corticospinal tract abnormalities are associated with weakness in multiple sclerosis. *AJNR Am J Neuroradiol*, *29*(2), 333-339. https://doi.org/10.3174/ajnr.A0788

Rowley C, L. I., Campbell JSW, Pike GB, Tardif CL. (2022). Acquisition optimization for cortical ihMTsat imaging. Proceedings of the International Society of Magnetic Resonance in Medicine, London.

Rowley, C. D., Campbell, J. S. W., Wu, Z., Leppert, I. R., Rudko, D. A., Pike, G. B., & Tardif, C. L. (2021). A model-based framework for correcting B1+ inhomogeneity effects in magnetization transfer saturation and inhomogeneous magnetization transfer saturation maps. *Magnetic Resonance in Medicine*, *86*(4). https://doi.org/10.1002/mrm.28831

Schiavi, S., Lu, P. J., Weigel, M., Lutti, A., Jones, D. K., Kappos, L., Granziera, C., & Daducci, A. (2022). Bundle myelin fraction (BMF) mapping of different white matter connections using microstructure informed tractography. *NeuroImage*, *249*, 118922. https://doi.org/10.1016/j.neuroimage.2022.118922

Schiavi, S., Ocampo-Pineda, M., Barakovic, M., Petit, L., Descoteaux, M., Thiran, J. P., & Daducci, A. (2020). A new method for accurate in vivo mapping of human brain connections using microstructural and anatomical information. *Science Advances*, *6*(31). https://doi.org/10.1126/sciadv.aba8245

Schiavi, S., Petracca, M., Battocchio, M., El Mendili, M. M., Paduri, S., Fleysher, L., Inglese, M., & Daducci, A. (2020). Sensory-motor network topology in multiple sclerosis: Structural connectivity analysis accounting for intrinsic density discrepancy. *Hum Brain Mapp*, *41*(11), 2951-2963. https://doi.org/10.1002/hbm.24989

Slater, D. A., Melie-Garcia, L., Preisig, M., Kherif, F., Lutti, A., & Draganski, B. (2019). Evolution of white matter tract microstructure across the life span. *Hum Brain Mapp*, *40*(7), 2252-2268. https://doi.org/10.1002/hbm.24522

Sled, J. G. (2018). Modelling and interpretation of magnetization transfer imaging in the brain. *NeuroImage*, *182*, 128-135. https://doi.org/10.1016/j.neuroimage.2017.11.065

Smith, R. E., Raffelt, D., Tournier, J. D., & Connelly, A. (2022). Quantitative streamlines tractography: methods and inter-subject normalisation. *Aperture Neuro*, *2*. https://doi.org/10.52294/ApertureNeuro.2022.2.NEOD9565

Stikov, N., Campbell, J. S., Stroh, T., Lavelee, M., Frey, S., Novek, J., Nuara, S., Ho, M. K., Bedell, B. J., Dougherty, R. F., Leppert, I. R., Boudreau, M., Narayanan, S., Duval, T., Cohen-Adad, J., Picard, P. A., Gasecka, A., Cote, D., & Pike, G. B. (2015). In vivo histology of the myelin g-ratio with magnetic resonance imaging. *NeuroImage*, *118*, 397-405. https://doi.org/10.1016/j.neuroimage.2015.05.023

Tournier, J. D., Smith, R., Raffelt, D., Tabbara, R., Dhollander, T., Pietsch, M., Christiaens, D., Jeurissen, B., Yeh, C. H., & Connelly, A. (2019). MRtrix3: A fast, flexible and open software framework for medical image processing and visualisation. *NeuroImage*, *202*, 116137. https://doi.org/10.1016/j.neuroimage.2019.116137

Tustison, N. J., Avants, B. B., Cook, P. A., Zheng, Y., Egan, A., Yushkevich, P. A., & Gee, J. C. (2010). N4ITK: improved N3 bias correction. *IEEE Trans Med Imaging*, *29*(6), 1310-1320. https://doi.org/10.1109/TMI.2010.2046908

van Gelderen, P., & Duyn, J. H. (2019). White matter intercompartmental water exchange rates determined from detailed modeling of the myelin sheath. *Magn Reson Med*, *81*(1), 628-638. https://doi.org/10.1002/mrm.27398

Varma, G., Girard, O. M., McHinda, S., Prevost, V. H., Grant, A. K., Duhamel, G., & Alsop, D. C. (2018). Low duty-cycle pulsed irradiation reduces magnetization transfer and increases the inhomogeneous magnetization transfer effect. *Journal of Magnetic Resonance*, *296*. https://doi.org/10.1016/j.jmr.2018.08.004

Varma, G., Girard, O. M., Prevost, V. H., Grant, A. K., Duhamel, G., & Alsop, D. C. (2017). In vivo measurement of a new source of contrast, the dipolar relaxation time, T(1D), using a modified inhomogeneous magnetization transfer (ihMT) sequence. *Magn Reson Med*, *78*(4), 1362-1372. https://doi.org/10.1002/mrm.26523

Veraart, J., Novikov, D. S., Christiaens, D., Ades-Aron, B., Sijbers, J., & Fieremans, E. (2016). Denoising of diffusion MRI using random matrix theory. *NeuroImage*, *142*, 394-406. https://doi.org/10.1016/j.neuroimage.2016.08.016

Veraart, J., Novikov, D. S., & Fieremans, E. (2018). TE dependent Diffusion Imaging (TEdDI) distinguishes between compartmental T2 relaxation times. *NeuroImage*, *182*, 360-369. https://doi.org/10.1016/j.neuroimage.2017.09.030

Wei, Y., Collin, G., Mandl, R. C. W., Cahn, W., Keunen, K., Schmidt, R., Kahn, R. S., & van den Heuvel, M. P. (2018). Cortical magnetization transfer abnormalities and connectome dysconnectivity in schizophrenia. *Schizophr Res*, *192*, 172-178. https://doi.org/10.1016/j.schres.2017.05.029

Yeatman, J. D., Dougherty, R. F., Myall, N. J., Wandell, B. A., & Feldman, H. M. (2012). Tract profiles of white matter properties: automating fiber-tract quantification. *PLoS ONE*, *7*(11), e49790. https://doi.org/10.1371/journal.pone.0049790

Zhang, F., Daducci, A., He, Y., Schiavi, S., Seguin, C., Smith, R. E., Yeh, C. H., Zhao, T., & O'Donnell, L. J. (2022). Quantitative mapping of the brain's structural connectivity using diffusion MRI tractography: A review. *NeuroImage*, *249*, 118870. https://doi.org/10.1016/j.neuroimage.2021.118870